\documentclass[pra,aps,twocolumn,10pt,showpacs,groupedaddress,superscriptaddress,floatfix]{revtex4-1}
\usepackage{amsmath,amsfonts,amssymb,graphics,graphicx,epsfig,color,times}
\usepackage[utf8x]{inputenc}
\usepackage{color}
\usepackage{bbm, dsfont} 
\usepackage{subfigure}
\usepackage{hyperref}
\usepackage{mathrsfs}
\usepackage{verbatim}
\begin{document}
\title{Cooperative single-photon subradiant states}
\author{H. H. Jen}
\email{sappyjen@gmail.com}
\affiliation{Institute of Physics, Academia Sinica, Taipei 11529, Taiwan}
\author{M.-S. Chang}
\affiliation{Institute of Atomic and Molecular Sciences, Academia Sinica, Taipei 10617, Taiwan}
\author{Y.-C. Chen}
\affiliation{Institute of Atomic and Molecular Sciences, Academia Sinica, Taipei 10617, Taiwan}

\date{\today}
\renewcommand{\r}{\mathbf{r}}
\newcommand{\f}{\mathbf{f}}
\renewcommand{\k}{\mathbf{k}}
\renewcommand{\r}{\mathbf{r}}
\def\bea{\begin{eqnarray}}
\def\eea{\end{eqnarray}}
\def\ba{\begin{array}}
\def\ea{\end{array}}
\def\bdm{\begin{displaymath}}
\def\edm{\end{displaymath}}
\def\red{\color{red}}
\begin{abstract}
We propose a set of subradiant states which can be prepared and detected in a one-dimensional optical lattice.\ We find that the decay rates are highly dependent on the spatial phases imprinted on the atomic chain, which gives systematic investigations of the subradiance in the fluorescence experiments.\ The time evolution of these states can have long decay time where up to hundred milliseconds of lifetime is predicted for one hundred atoms.\ They can also show decayed Rabi-like oscillations with a beating frequency determined by the difference of cooperative Lamb shift in the subspace.\ Experimental requirements are also discussed for practical implementation of the subradiant states.\ Our proposal provides a novel scheme for quantum storage of photons in arrays of two-level atoms through the preparation and detection of subradiant states, which offer opportunities for quantum many-body state preparation and quantum information processing in optical lattices.
\end{abstract}
\pacs{42.50.Gy, 42.50.Ct, 42.50.Dv}
\maketitle
\section{Introduction} 
Cooperative light-matter interaction \cite{Mandel1995} plays an important role in the aspects of correlated spontaneous emissions, quantum entanglement \cite{Hammerer2010}, and preparation of exotic quantum states \cite{Hartmann2006, Greentree2006, Lewenstein2007, Chang2008, Georgescu2014}.\ Superradiance \cite{Dicke1954, Gross1982} and associated cooperative Lamb shift (CLS) \cite{Friedberg1973, Scully2009} are examples that demonstrate the collectivity in exchanging photons within an atomic ensemble.\ Subradiance, on the other hand, is a less aware collective effect complementary to the superradiance, which spontaneously emits light in a rate lower than the natural one.\ Single-photon superradiance \cite{Scully2006, Eberly2006, Mazets2007} and subradiance are of great interests in that the dimensionality of the Hilbert space is restricted to the order of the number of atoms $N$ only, thus simplifies the dynamical couplings with each other.\ 

Resonant dipole-dipole interaction \cite{Stephen1964, Lehmberg1970} is regarded to initiate the superradiance with properties of directional emissions in an enhanced spontaneous decay rate.\ Because of its spatial dependence in the short range ($\propto$ $1/r^3$) and the long range ($\propto$ $1/r$), the decay behavior heavily depends on the density and geometry of the interacting medium, and additionally excitation polarizations.\ Recent experiments demonstrated a shorter timescale of the second-order correlation of two photons in the cascade atomic ensembles \cite{Chaneliere2006, Srivathsan2013}, and a significant redshift of CLS in various atomic systems including the embedded Fe atoms in the planar cavity \cite{Rohlsberger2010}, an atomic vapor \cite{Keaveney2012}, an ionic atomic array \cite{Meir2014}, and a cold atomic ensemble \cite{Pellegrino2014}.\ Subradiant lifetime was also recently measured in plasmonic ring nanocavities \cite{Sonnefraud2010}, ultracold molecules \cite{McGuyer2015}, and a cold atomic ensemble \cite{Guerin2016}.

In the next section, we investigate the construction of singly-excited states which are candidates for superradiance and subradiance emerged from the resonant dipole-dipole interactions.\ We introduce the cooperative single-photon subradiant states that can be prepared in our proposed scheme by imprinting the required phases via a pulsed gradient magnetic or electric fields.\ In Sec. III, we discuss the experimental realizations of the subradiant states, and we show the time evolutions of these subradiant states that can be measured in fluorescence experiments in Sec. IV.\ Finally we address the influence of the phase imperfections and conclude in Sec. V. 

\section{Construction of cooperative singly-excited states}
In a setting where a (near) resonant single photon is driving a system of two-level atoms ($|g\rangle$ and $|e\rangle$ for the atomic ground and excited sates respectively), the superradiance and accompanied CLS \cite{Scully2009, Jen2015} emerge from the symmetrical and singly-excited state or timed Dicke state \cite{Scully2006},
\bea
|\phi_N\rangle=\frac{1}{\sqrt{N}}\sum_{\mu=1}^N e^{i\k\cdot \r_\mu}|e\rangle_\mu|g\rangle^{\otimes(N-1)},
\eea
where $\k$ is the wavevector of the excitation pulse of a single photon, and the tensor product denotes the rest of ($N-1$) ground state atoms other than the one being excited, that is $|e\rangle_\mu|g\rangle^{\otimes(N-1)}$ $\equiv$ $|\psi_\mu\rangle$.\ Since the symmetrical state is approximately decoupled from the rest of ($N-1$) orthogonal and nonsymmetrical (NS) states $\{|\phi_{m\neq N}\rangle\}$, the emission long after the superradiance (afterglow if there is any because of extremely low probability in the limit of extended medium) suggests the occupation of the these states \cite{Mazets2007}.\ These NS states responsible for the spontaneous emission in a rate lower than the natural decay one are subradiant states, which were observed in a large cloud of cold atoms \cite{Guerin2016}.\ A recent proposal also suggested for preparation and detection of subradiant states in the scheme based on subensembles \cite{Scully2015}.\ 

A complete Hilbert space for a single photon interacting with the atomic ensemble has been proposed for these normalized NS states \cite{Mazets2007},
\bea
|\phi_m\rangle&=&\sum_{\mu=1}^{N}\Big(C-\frac{\sqrt{N}+1}{N-1}\delta_{\mu N}-\delta_{\mu m}\Big)e^{i\k\cdot\r_\mu}|\psi_\mu\rangle,
\eea
for $m$ $\in$ [$1$ , $N-1$] with $C$ $\equiv$ $(1+1/\sqrt{N})/(N-1)$.\ An alternative construction of the complete space is proposed in terms of Dicke's states where the formalism of singlet and triplet states in the angular momentum eigenstates provides the foundation to construct the Hilbert space \cite{Svidzinsky2008},
\bea
|\phi_m\rangle=\frac{\sum_{\mu=1}^m e^{i\k\cdot\r_\mu}|\psi_\mu\rangle-me^{i\k\cdot\r_{m+1}}|\psi_{m+1}\rangle}{\sqrt{N(N-1)}}.
\eea
The difference of the above two NS states is that the first involves all the atoms whereas the second involves a subset of atoms in specific positions.\ In the perspective of experimental preparation of those NS states, both require exquisite control on the coefficients of the constituent atoms, including the amplitudes and phases.\ An issue associated with that is those analytic NS states are not the eigenstates of the Hamiltonian, and separately assessing each NS state through global control is difficult to achieve.\ As such it may require complicated individual addressing fields with sub-wavelength resolution in order to independently manipulate each and every constituent atom \cite{Plankensteiner2015}.\ Alternatively subradiant states can be prepared in the setting of spatially-separated subensembles \cite{Scully2015}, in which preparation of specific subradiant states is more feasible, though more resources are required for more subensembles.

Below we propose a complete space that involves all the atoms collectively, which can be prepared in the one-dimensional (1D) optical lattice (OL) with one atom per site.\ The states are prepared by a pulsed global control field following a single-photon excitation with equal probability on every atom, thus greatly improves their practical accessibility experimentally. 

\subsection{De Moivre states} 
To construct the complete N states for single-photon excitation, we exploit De Moivre's formula which is originally used for the nth root of unity, that is for a complex number $z$ such that $z^n$ $=$ $1$.\ We take the roots of $z^N$ $=$ $1$ as indications of the coefficients for the so-called De Moivre (DM) states we propose in the below, which are $e^{i2\pi m/N}$ for $m$ $\in$ [$1$ , $N$].\ The identity of the sum of all roots, SR($n$) $=$ $0$ for all $n$ except for $n$ $=$ $1$, indicates the orthogonality of the DM states.\ We construct the complete Hilbert space with the DM states including the symmetric one $|\phi_N\rangle$ \cite{Vetter2016,note1},
\bea
|\phi_m\rangle=\sum_{\mu=1}^N \frac{e^{i\k\cdot\r_\mu}}{\sqrt{N}}e^{i\frac{2m\pi}{N}(\mu-1)}|\psi_\mu\rangle,
\eea
where the normalization is ensured  with the orthonormality $\langle\phi_m|\phi_n\rangle$ $=$ $\sum_{\mu=1}^Ne^{i\frac{2\pi}{N}(\mu-1)(m-n)}/N$ $=$ $\delta_{m,n}$.\ In each DM state the phases of the atoms in the 1D OL increase discretely and linearly while the amplitudes of the coefficients are equal.\ This contrasts the DM states with the aforementioned NS states where both the amplitudes and the phases of their coefficients need to be specifically constructed.\ The DM states also remind us of the discrete Fourier transform in that their coefficients are exactly the Fourier series bases since the coefficients $e^{i2\pi m(\mu-1)/N}$ are $N$-periodic.\ We note that the DM states introduced here are correlated ones induced by the long-ranged dipole-dipole interaction, which is responsible for the cooperative spontaneous emissions, and they can be prepared in a collective way by an external field gradient.

\begin{figure}[t]
\centering
\includegraphics[width=8.5cm,height=6.5cm]{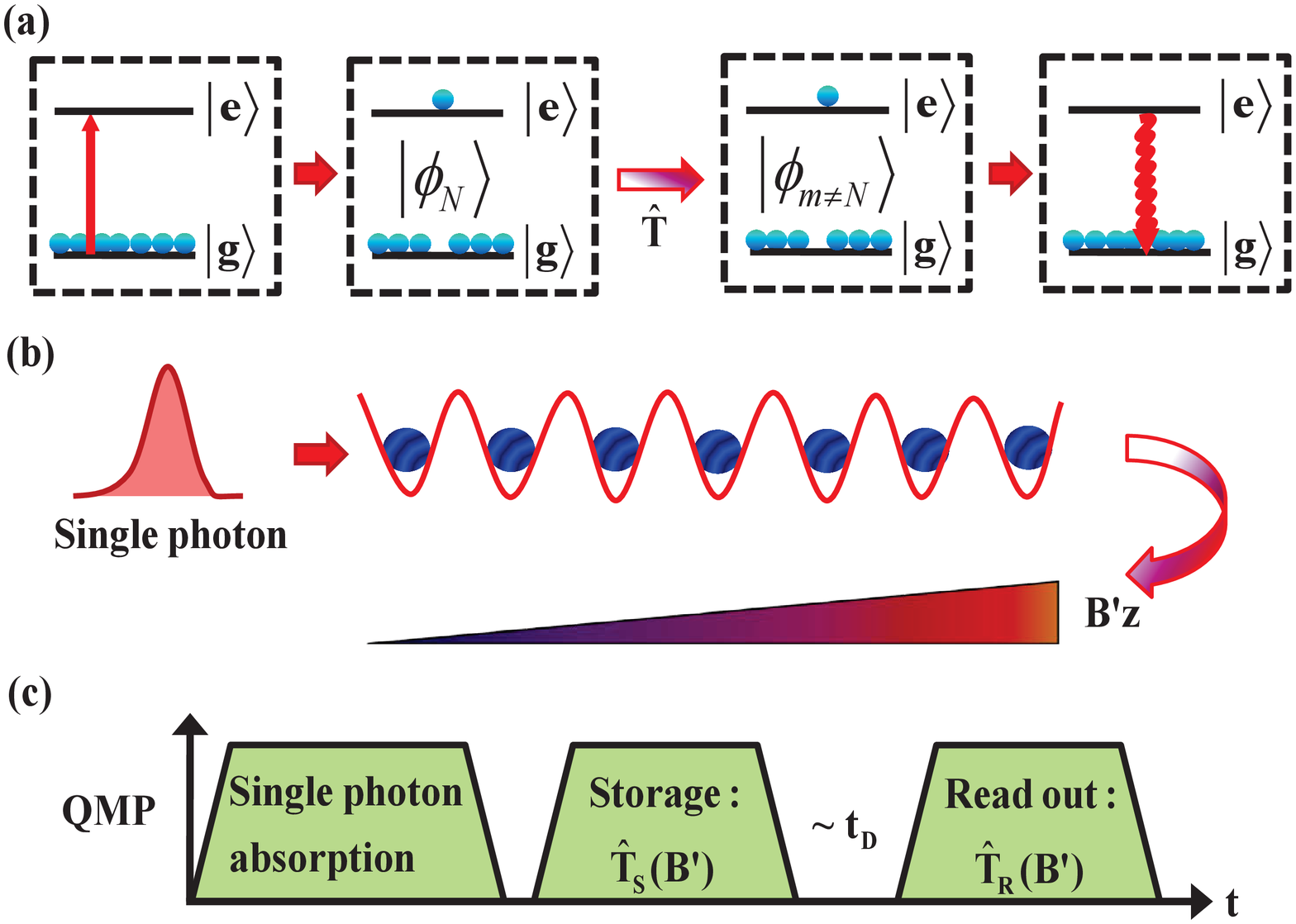}
\caption{(Color online) Preparation of De Moivre states and quantum memory protocol (QMP).\ (a) Creation of DM states from the symmetrical $|\phi_N\rangle$ evolved by a unitary transformation $\hat{T}$.\ (b) Schematic demonstration of single photon absorption by 1D atomic array to form $|\phi_N\rangle$ followed by a pulsed magnetic field gradient to imprint the necessary phases of the DM states.\ (c) Quantum memory protocol for quantum storage of single photon.\ After single photon absorption, the system is excited to $|\phi_N\rangle$ as in (a), and the subradiant state is prepared and evolved from $|\phi_N\rangle$ by $\hat{T}_S(B')$ for quantum storage with a time delay $t_D$.\ The storage protocol reads out the photon by $\hat{T}_R(B')$ $=$ $\hat{T}^{-1}_S(B')$.}\label{fig1}
\end{figure}

\section{Experimental realizations} 
To illustrate, we first propose to use a magnetic field gradient to introduce the linearly increasing phases along the 1D OL shown in Fig. \ref{fig1}.\ The system can be first prepared in $|\phi_N\rangle$ by absorbing a single photon.\ Then a gradient Zeeman field is pulsed on to evolve $|\phi_N\rangle$ to $|\phi_{m\neq N}\rangle$.\ This evolution can be described by a unitary transformation $\hat{T}$ $\equiv$ $e^{iV_B\tau_B}$, where the interaction energy $V_B$ $=$ $-\boldsymbol{\mu}\cdot\mathbf{B}$, and $\tau_B$ is the interaction time.\ After the evolution the spatial phases of $V_B\tau_B$ $=$ $m_Fg_F\mu_B B'z\tau_B$ are imprinted on the atoms in the $m_F$ Zeeman sublevel of the $F$ hyperfine state.\ Here $g_F$ is the Land\'{e} $g$ factor, $\mu_B$ is the Bohr magneton, $B'$ is field gradient, and $z$ represents the positions of the atoms.\ The measurement of fluorescence afterwards verifies the state we prepare, which therefore provides a systematic way to study super- and sub-radiance of the cooperative single-photon DM states.\ The subradiant state can be made to store light quanta for much longer time, thus is potentially useful for quantum memory of light \cite{Afzelius2015}.

We use the D$_2$ transition of $^{87}$Rb atoms as an example, and we choose the ground and excited states as $|5^2$S$_{1/2}$, $F=2$, $m_F=2\rangle$ with $g_F$ $=$ $1/2$ and $|5^2$P$_{3/2}$, $F'=3$, $m_F=3\rangle$ with $g_F$ $=$ $2/3$, respectively.\ The required phase difference for two adjacent sites is $2m\pi/N$ for the $m$th DM states, with $m$ $=$ $1,2,...,N$.\ Thus to access the whole Hilbert space of the DM states, the maximum phase difference needs to be as large as $2\pi$.\ Therefore $B'$ should be as large as $0.92/(d_s'\tau_B')$ in unit of [mG$/\mu$m], with the pulse duration $\tau_B'$ in milliseconds and the lattice spacing $d_s'$ in $\lambda$ which is $780$ nm for the D$_2$ transition.\ For the order of $d_s'$ $=$ $1$ we consider here, the field gradient would be around $92$ mG$/\mu$m for $10$ microseconds of interaction time, which is within reach of typical experiments.\ In the perspective of preparing a specific $m$th DM state, a longer atomic chain of larger $N$ also requires a less field gradient or a shorter interaction time to imprint the phase gradients across the whole chain.\ For example of $N$ $=$ $500$ we will demonstrate later, the field gradient just requires $1.8$ mG$/\mu$m or the interaction time becomes $200$ nanoseconds to prepare $m$ $=$ $10$ DM state which still has a reduction of decay rate in the order of $10^{-2}\Gamma$.

An issue of QMP in Fig. \ref{fig1}(c), however, is the lack of efficiency due to the creation of the superradiant state in the beginning of the DM state preparation, which decays faster than the free space decay rate ($\sim$ $26$ ns).\ The efficiency can be estimated as $e^{-\Gamma_N\tau_B}$ where $\Gamma_N$ $\gtrsim$ $1/26$ ns$^{-1}$ is the decay rate of the superradiant state we initially prepare.\ To get around this low efficiency, we can utilize linear Stark shift to imprint the required phases in a rate of $10$'s MHz to restore the efficiency in QMP.\ Both continuous wave and ultrashort light pulse can be used to fast control the phases with higher efficiency, and a sidewise single-photon excitation with $\pi$ phase shifters switches between super- and sub-radiant states in an ultrashort time scale was proposed \cite{Scully2015}.\ The ac Stark shift can be calculated as $U(\omega,I)$ $=$ $-\alpha_{D_2}(\omega)E^2/2$ $=$ $-\alpha_{D_2}(\omega)I/(2\epsilon_0c)$, where $\alpha_{D_2}(\omega)$ $=$ $\alpha_{D_2}(0)\omega_{eg}^2/(\omega_{eg}^2-\omega^2)$ \cite{Grimm2000, Sparkes2010, Steck2015}, in the limit of large detuning from the D$_2$ transition by using a laser field, where rotating wave approximation is still valid.\ Here $\alpha_{D_2}(0)$ is the static atom polarizability \cite{Steck2015}.\ On preparing $m$th DM state, we then estimate the required maximum ac Stark field intensity as $I$ $\sim$ $2.6\times 10^{12}$ [mW/cm$^2$] with an interaction time of $1$ ns and $m$ $\leq$ $100$.\ This can be readily achieved by a typical pulsed laser.\ For example, given a pulsed laser with an average power of $200$ mW, a repetition rate of $1$ kHz, and a pulse width of $1$ ns, when it is focused to a waist of $50$ $\mu$m, its peak intensity is as high as $I_p$ $=$ $2.6\times 10^{12}$ mW/cm$^2$.\ With a continuous wave laser at a much smaller detuning, say $\Delta_{3/2,F=3}$ $=$ $\omega_{eg}-\omega$ $\approx$ $100\Gamma$, then the linear Stark field gradient requires an intensity of $I$ $\approx$ $1.26\times 10^6$ mW/cm$^2$ for $1$ ns interaction time for preparing the $m$ $=$ $7$ DM state.\ This can be achieved by focusing a $100$ mW laser to a beam waist of $50$ $\mu$m.\ Irrespective of the recipes introduced above for phase imprinting, with heralded single photon source and post selection, the distilled write efficiency should be greatly restored in the preparation of the DM states for applications in many-body physics and quantum memory. 

The technique of position-dependent phase imprinting \cite{Kraus2006} reminds us of controlled reversible inhomogeneous broadening (CRIB) that is used to implement the quantum memory of light in two-level atoms of praseodymium dopants in yttrium orthosilicate (Pr$^{3+}$:Y$_2$SiO$_5$) \cite{Hetet2008, Hedges2010} or warm $\Lambda$-type rubidium vapour \cite{Hosseini2011}, where CRIB is introduced for efficient absorption of light.\ In the case here, the inhomogeneous field is to unitarily introduce the spatial phases to the atoms after absorption of the light quanta.\ Readout process in the proposed protocols is straightforward in that the specifically prepared DM states are made to evolve back to $|\phi_N\rangle$, which is a superradiant readout.

\begin{figure}[t]
\centering
\includegraphics[width=8.5cm,height=4.5cm]{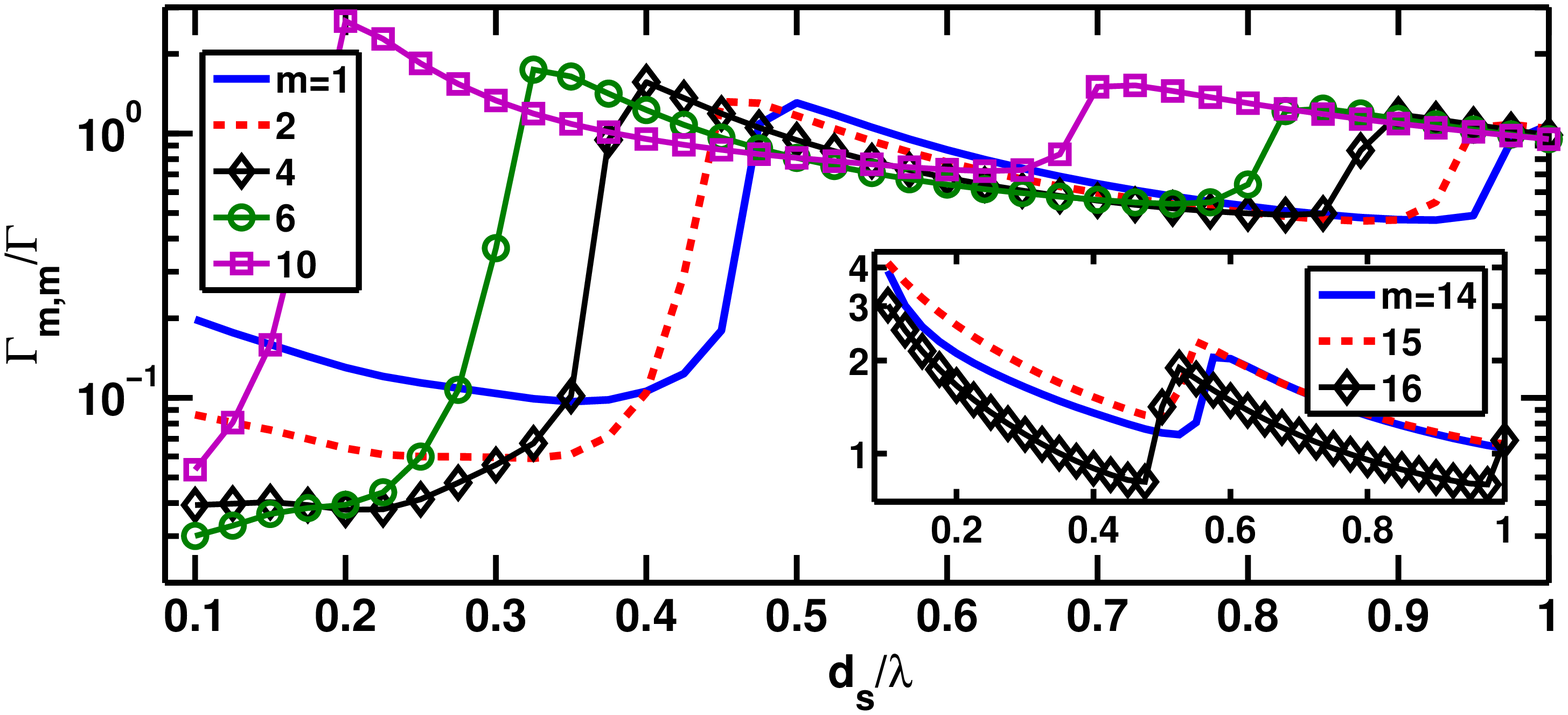}
\caption{(Color online) The coupling strengths $\Gamma_{m,m}$ of the DM states for various lattice spacings with $N$ $=$ $16$.\ Small coupling strengths appear mostly with $m$ $\lesssim$ $N/2$ at small lattice spacing $d_s$ $\approx$ $0.1-0.4\lambda$, while relatively large ones are demonstrated in the inset near $m$ $\approx$ $N$.}\label{fig2}
\end{figure}

\section{Time evolution of DM states} 
To investigate what are the small subradiant eigenvalues assessed by the subradiant DM states and what is the optimal lattice spacing for the lowest eigenvalues, we first study the coupling strength of the resonant dipole-dipole interaction in the DM state bases.\ Using linear polarization of light propagating along the OL (that is $\hat{d}$ $\perp$ $\hat{r}_{\mu\nu}$), we define the coupling strength as 
\bea
\Gamma_{m,m}=-2{\rm Re}\bigg[\bigg\langle\phi_m\bigg|\sum_{\mu,\nu}M_{\mu\nu}\bigg|\phi_m\bigg\rangle\bigg],
\eea 
where $M_{\mu\nu}$ $\equiv$ $\frac{\Gamma}{2}(-F_{\mu\nu}+i2G_{\mu\nu}\delta_{\nu\neq\mu})$, and $F_{\mu\nu}$, $G_{\mu\nu}$ are decay rates and energy shifts from the resonant dipole-dipole interaction between any pair of atoms (see Appendix A for definitions).\ In Fig. \ref{fig2} we show the coupling strengths for the DM states, and find that small coupling strengths mostly lie at lattice spacings less than $0.4\lambda$.\ The period of $\lambda/2$ in $d_s$ for the coupling strengths is due to the sinusoidal functions in the dipole-dipole interaction, which is equivalent to a period of $\pi$.\ Similar periodic coupling strengths are demonstrated in infinite atomic lattices with a period of $2\pi$ \cite{Nienhuisi1987}.\ The coupling strengths saturate quickly as $N$ increases, therefore here we show for the case of $N$ $=$ $16$ without loss of generality.\ Large coupling strengths can also be seen in the inset, showing continuously shifted local peaks from small to large $m$ states due to equidistant change of $m$.\ From the coupling strength calculations, we later choose specifically $d_s$ $=$ $0.1$ and $0.25\lambda$ that correspond to the lowest coupling strengths for the DM states of $m$ $=$ $6$ and $4$ respectively.\ These specific DM states will be shown later to demonstrate the smallest decay rates (longest lifetimes).

Next we investigate the time evolution of the DM states which can be observable in the fluorescence experiments.\ Since the resonant dipole-dipole interaction is long ranged, there is no exact analytical form of the eigenstates in a finite 1D OL.\ Therefore the DM states we prepare in the 1D OL would simultaneously couple to several eigenstates that we numerically solved.\ In Fig. \ref{fig3} we plot the normalized weightings (projections) defined in Appendix B for DM states.\ The weightings indicate how significantly the eigenvalues $\lambda_n$'s contribute to the time evolution of the DM states, and generally they are highly localized within two to three eigenstates.\ This means that the DM states form nearly closed subspaces for the original eigenstates.\ For instance in (a) we can see that $m$ $=$ $2$ and $11$ DM states form an almost closed subspace for $n$ $=$ $9$ and $10$ eigenstates with $98\%$ weightings.\ This is reflected by that our DM states resemble a decayed Rabi oscillation with a beating frequency at the difference of CLS's for the eigenstates spanning the subspace.\ The physical origin of the localized subspaces is likely due to the effect of finite atomic chain.\ We find that the normalized weightings of some of the specific DM states (for example the state with the longest lifetime for $d_s$ $=$ $0.1$ and $0.25\lambda$) approach unity as the number of atoms increases, which suggests that these DM states are almost identical to the eigenstates in a longer atomic chain.\ In finite atomic array a DM state generally still projects to a small number of nearby eigenstates. 

\begin{figure}[t]
\centering
\includegraphics[width=8.5cm,height=4.5cm]{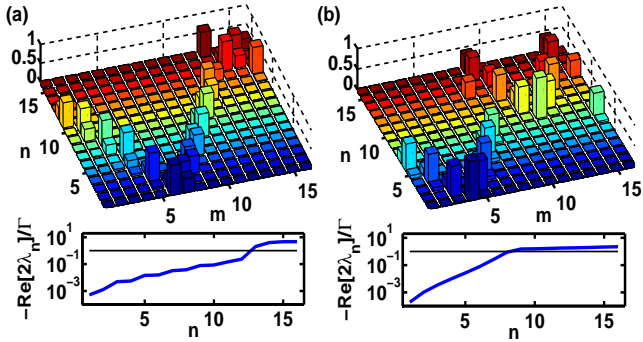}
\caption{(Color online) Normalized weightings (projections) of DM states $|\phi_m\rangle$ on the eigenstates $|\phi'_n\rangle$ for $N$ $=$ $16$.\ The weightings are observed in localized groups of the eigenstates for both (a) $d_s$ $=$ $0.1\lambda$ and (b) $0.25\lambda$.\ The corresponding ascending order of the real part of the eigenvalues indicates the distribution of the superradiant (above $\Gamma$) and subradiant (below $\Gamma$) decay constants.\ A horizontal line is used to guide the eye for a natural decay constant which is $-$Re$[2\lambda_n]/\Gamma$ $=$ $1$.}\label{fig3}
\end{figure}

In Fig. \ref{fig3}(a) for a small lattice spacing of $d_s$ $=$ $0.1\lambda$, we clearly see groups of subradiant and superradiant eigenvalues in ascending order.\ Specifically for $d_s$ $=$ $0.25\lambda$ in (b), more superradiant ones are lying closer to $\lambda_n$ $=$ $\Gamma/2$, showing less enhanced decay rates due to the larger lattice spacing.\ We note that the nonsymmetric DM states ($m$ $\neq$ $N$) can be superradiant or subradiant while the one with $m$ $=$ $N$ is always superradiant.\ For even larger $d_s$, we have more eigenvalues distributing near $\Gamma/2$ as expected for the system approaching the limit of independent atoms.

\begin{figure}[t]
\centering
\includegraphics[width=8.5cm,height=4.5cm]{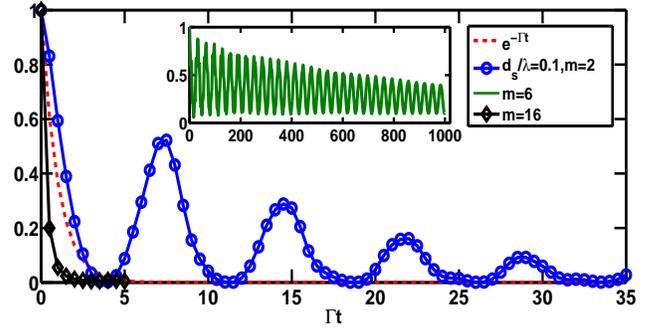}
\caption{(Color online) Time evolutions of cooperative single-photon subradiant states.\ For $N$ $=$ $16$ and $d_s$ $=$ $0.1\lambda$, we demonstrate the time evolutions of DM states of $m$ $=$ $2$ ($\circ$, oscillatory and subradiant), $m$ $=$ $6$ (inset, oscillatory and subradiant), and $m$ $=$ $16$ ($\diamond$, superradiant), compared to the natural decay one $e^{-\Gamma t}$ (dash).}\label{fig4}
\end{figure}

In Fig. \ref{fig4}, we plot the time evolution of the DM state probabilities specifically for $m$ $=$ $2$ and $6$ and compare that with the case of the independent atoms $(\propto e^{-\Gamma t})$.\ The DM state of $m$ $=$ $2$ shows Rabi-like oscillation, which is due to that it forms a subspace with the $m$ $=$ $11$ DM state.\ The beating frequency in the plot is estimated as $0.863\Gamma$, which has less than $1.2\%$ relative error compared to $0.853\Gamma$, the difference of CLS's of $n$ $=$ $9$ and $10$.\ We note that the $1.2\%$ error on the beating frequency is due to the $2\%$ projection (weightings) of the $m$ $=$ $11$ DM state on other eigenstates.\ In this way the DM states provide a systematic measurement of relative CLS from the fluorescence experiments.\ The symmetric DM state $|\phi_N\rangle$ with $m$ $=$ $16$ exhibits superradiance as expected, which however involves several superradiant eigenmodes as seen in Fig. \ref{fig3}(a).\ A very interesting and important observation is found for the DM state of $m$ $=$ $6$ where we plot the envelopes of the long-time emission intensity.\ This DM state is mostly comprised of two eigenmodes with the lowest real parts of the two eigenvalues ($5.1\times 10^{-4}$ and $1.3\times 10^{-3}\Gamma$) with $99\%$ weightings among all. 

\begin{figure}[t]
\centering
\includegraphics[width=8.5cm,height=4.5cm]{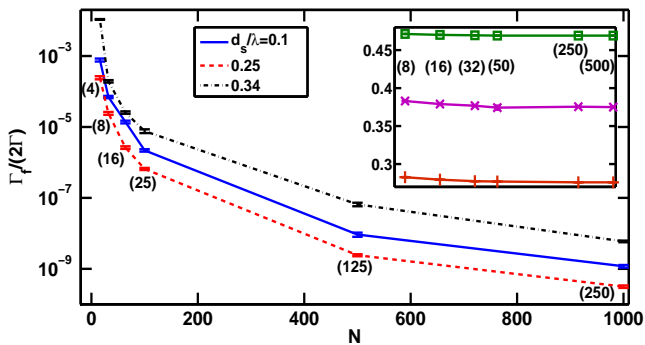}
\caption{(Color online) Fitted decay constants.\ For atom numbers $N$ $=$ $16$, $32$, $64$, $100$, $500$, and $1000$, we show the fitted lowest decay constant $\Gamma_f$ in logarithmic scale by the envelope of $e^{-\Gamma_f t}$ for $d_s$ $=$ $0.1$ (solid), $0.25$ (dash), $0.34\lambda$ (dash-dot), and for $d_s$ $=$ $0.5$ ($\times$), $0.68$ ($+$), and $1\lambda$ ($\square$) in the inset where the cases of $0.5$ and $1\lambda$ share the same DM states denoted by the parenthesis.\ The parenthesis $(m)$ shows the $m$th DM state which exhibits the lowest $\Gamma_f$ for the cases of $d_s$ $=$ $0.25$ and $0.5\lambda$ for demonstration.\ The error bar indicates the deviation from the fitted exponential decay constants with or without the intercept of the zero time point.}\label{fig5}
\end{figure}

To determine the lowest decay rates of the subradiant DM states, the emission envelopes are fit to an exponential decay, $e^{-\Gamma_f t}$, for different numbers of atoms and lattice spacings shown in Fig. \ref{fig5}.\ We show that decay rates as low as the order of $10^{-7}\Gamma$ is attainable for $100$ atoms, which is equivalent to the lifetime of several hundred milliseconds for rubidium atoms.\ When $d_s$ $\gtrsim$ $0.5\lambda$ as shown in the inset, the decay constant is shown barely dependent on the atom numbers.\ Each parenthesis shows the specific $m$th DM state mostly occupying the eigenmodes with the lowest real part of the eigenvalue, which scales with atom numbers.\ The cases of the lattice spacings at $0.5$ and $1\lambda$ sharing the same DM states again reflects the period of $\lambda/2$ in the coupling strengths.\ For even longer 1D chains, the lifetime can be longer and seems to approach indefinitely to zero decay for $d_s$ shorter than $0.4\lambda$, although it is more challenging for the experiments in the perspectives of stability and controllability. 

Our proposal for the DM states implemented in a 1D OL can potentially be realized in 1D hard core bosons \cite{Paredes2004, Kinoshita2004}, ions in a linear Paul trap \cite{Meir2014}, color center defects in diamond \cite{Weimer2013, Sipahigil2014}, or atom-fiber system \cite{Chang2012}.\ With a scalable 1D array, decoherence-free regime can be feasible, which adds the richness and robustness to quantum information network.\ Our setting is also alternative to investigating the many-body long-range interactions in the alkaline-earth-metal atoms \cite{Olmos2013} and the cooperative behavior in the square and kagome lattices \cite{Bettles2015}.\ For preparation of the single-photon subradiant states, a scheme with Rydberg atomic excitations \cite{Saffman2010, Peyronel2012} is sufficient to generate singly-excited cooperative states via dipole blockade effects along with phase imprinting.\ We expect even richer dynamical couplings between atoms by introducing this additional long-ranged dipole-dipole interactions.

\section{Discussion and conclusion}
Preparation of the DM states relies on the proper phase imprinting on the atoms, which is achieved by controlling the interaction time of the atoms with the applied field gradient.\ To investigate the influence of the imprecision of phase imprinting on the fidelity of the DM states due to either inaccurate field strength or interaction time (or both), we compare the changes in the couplings to the eigenstates and in fluorescence measurements with phase errors of $0.2\pi$ and $0.4\pi$ across the whole atomic chain.\ We find that generally there are two-fold changes, i.e. the projection to the major subset of the eigenstates and the projections to the constituent eigenstates within this subset, which both contribute to the infidelity of the DM state preparation.\ The first change indicates a coupling outside the localized subset of the eigenstates, while both influence the measurement of DM states lifetimes.  

For the case of $N$ $=$ $16$, $d_s$ $=$ $0.1\lambda$, and $m$ $=$ $2$ DM state, with a $0.2\pi$ $(0.4\pi)$ phase offset across the atomic chain, we have equivalently $5\%$ $(10\%)$ of phase error, and the projection to the major subset is reduced to $92\%$ $(82\%)$ compared to $98\%$ in the perfect case.\ The corresponding error in measuring the beating frequency from cooperative Lamb shifts raises to $1.2\%$ $(5\%)$, respectively, compared to $1.2\%$ without a phase error.\ The fitted decay constant $\Gamma_f/(2\Gamma)$ lowers to $0.0725\pm 0.0015$ $(0.0665\pm 0.0015)$ compared to $0.078\pm 0.001$ without a phase error.\ We further find that the effect of the imprecise phase imprinting is more prominent for more atoms.\ With $N$ $=$ $64$ and $m$ $=$ $2$ DM state, we find the projection to the major subset reduces to $87\%$ $(71\%)$ with a phase error of $0.2\pi$ $(0.4\pi)$ compared with $94\%$ in the perfect case.\ Also in this case the fitted decay constant raises to $0.018\pm 0.002$ $(0.019\pm 0.003)$ compared to $0.017\pm 0.001$ in the perfect case.\ Similar results are found for larger lattice spacings.\ For $d_s$ $=$ $0.4\lambda$ and a phase offset of $0.2\pi$ $(0.4\pi)$ given $N$ $=$ $16$ and $m$ $=$ $2$ DM state, we find that the projection to the major subset remains as high as $98\%$ $(98\%)$ compared to $99\%$ in the perfect case, while the projections to the constituent eigenstates inside the major subset changes by $28\%$ $(50\%)$.\ The fitted decay constant then raises to $0.0195\pm 0.0035$ $(0.0215\pm 0.0045)$ compared to $0.0175\pm 0.0025$ in the perfect case.\ Experimentally a phase error at few percent level should be achievable relatively easily, and as such the fidelity of the proposed DM states and the associated lifetimes should not deviate much from the perfect case.

In conclusion, we propose a complete Hilbert space of cooperative single-photon states that can be prepared and manipulated in an array of two-level atoms trapped in a one-dimensional optical lattice.\ Those cooperative subradiant states can be systematically studied by varying the spatially-increased phases we imprint on the atoms utilizing either a Zeeman or Stark field gradient pulse.\ Cooperative Lamb shift can also be studied in this setting by fluorescence experiments.\ Hundred milliseconds of lifetime can be observable in several tens of atoms, serving a potentially robust quantum memory of light.

\section*{Acknowledgements}
This work is supported by the Ministry of Science and Technology, Taiwan, under Grant No. MOST-101-2112-M-001-021-MY3 and MOST-103-2112-M-001-011.\ We are also grateful for the support of NCTS and the fruitful discussions with S.-Y. Lan.
\appendix
\section{Resonant dipole-dipole interaction}
The theoretical analysis is based on the Hamiltonian ($V_{\rm I}$) of a quantized radiation field interacting with a two-level atomic ensemble of N atoms.\ In interaction picture we have ($\hbar$ $=$ $1$)
\bea
V_{\rm I}=-\sum_{\mu=1}^{ N}\sum_{\k,\lambda} g_{\k}(\epsilon_{\k,\lambda}\cdot\hat{d})\hat{S}_\mu[e^{-i(w_{\k}t-\k\cdot\r_\mu)}\hat{a}_{\k,\lambda}+\rm h. c.] \label{V},\nonumber\\
\eea
and the dipole operator is defined as 
\bea
\hat{S}_\mu\equiv\hat{\sigma}_\mu e^{-i\omega_{eg}t}+\hat{\sigma}_\mu^\dag e^{i\omega_{eg}t},
\eea
where the lowering operator is $\hat{\sigma}_\mu$ $\equiv$ $|g\rangle_\mu\langle e|$ with the transition frequency $\omega_{eg}$ $=$ $\omega_e$ $-$ $\omega_g$.\ The coupling coefficient is $g_{\k}$ $\equiv$ $(g||\hat{d}||e)\mathcal{E}(k)$ where the matrix element of the dipole moment $\hat{d}$ is independent of the hyperfine structure, and $\mathcal{E}(k)$ $=$ $\sqrt{kc/(2\epsilon_0 V)}$ with the quantization volume $V$.\ Photon polarization is $\epsilon_{\k,\lambda}$, and the unit vector of the dipole operator is $\hat{d}$.\ The bosonic photon operators should also satisfy the commutation relation $[\hat{a}_{\k,\lambda},\hat{a}^\dag_{\k',\lambda'}]$ $=$ $\delta_{\k,\k'}\delta_{\lambda,\lambda'}$.\ We use the dipole approximation for the Hamiltonian and keep the counter rotating-wave parts to correctly account for the energy shift of the dipole-dipole interaction in the below.\   

To solve Heisenberg equations of motion for the above Hamiltonian, we derive the effective coupled equations for arbitrary atomic operators $\hat{Q}$ in a Lindblad form that
\bea
\frac{\dot{\hat{Q}}}{\Gamma}&=&i\sum_{\mu\neq\nu}^{ N}\sum_{\nu=1}^{ N} G_{\mu\nu}\left[\hat{\sigma}_{\mu}^\dag\hat{\sigma}_\nu,\hat{Q}\right]\nonumber\\
&+&\sum_{\mu=1}^{ N}\sum_{\nu=1}^{ N} F_{\mu\nu}\left[\hat{\sigma}_{\mu}^\dag\hat{Q}\hat{\sigma}_\nu-\frac{1}{2}\left(\hat{\sigma}_{\mu}^\dag\hat{\sigma}_\nu\hat{Q}
+\hat{Q}\hat{\sigma}_{\mu}^\dag\hat{\sigma}_\nu\right)\right],
\eea
where $F_{\alpha,\beta}$ and $G_{\alpha,\beta}$ are defined as \cite{Lehmberg1970}
\bea
F_{\mu\nu}(\xi)&\equiv&
\frac{3}{2}\bigg\{\left[1-(\hat{d}\cdot\hat{r}_{\mu\nu})^2\right]\frac{\sin\xi}{\xi}\nonumber\\
&+&\left[1-3(\hat{d}\cdot\hat{r}_{\mu\nu})^2\right]\left(\frac{\cos\xi}{\xi^2}-\frac{\sin\xi}{\xi^3}\right)\bigg\},\\
G_{\mu\nu}(\xi)&\equiv&\frac{3}{4}\bigg\{-\Big[1-(\hat{d}\cdot\hat{r}_{\mu\nu})^2\Big]\frac{\cos\xi}{\xi}\nonumber\\
&+&\Big[1-3(\hat{d}\cdot\hat{r}_{\mu\nu})^2\Big]
\left(\frac{\sin\xi}{\xi^2}+\frac{\cos\xi}{\xi^3}\right)\bigg\}.
\eea
Here $\xi$ $=$ $|\mathbf{k}| r_{\mu\nu}$, and $r_{\mu\nu}$ $=$ $|\mathbf{r}_\mu-\mathbf{r}_\nu|$ with the transition wave number $|\mathbf{k}|$.\ This is the origin of resonant dipole-dipole interaction induced by the common light-matter interaction.\ $F_{\mu\nu}$ and $G_{\mu\nu}$ are spatially-dependent decay rates and cooperative Lamb shifts respectively, where $F_{\mu\nu}$ approaches $1$ as $\xi\rightarrow 0$ while $G_{\mu\nu}$ becomes divergent.\ The divergence means the inaccuracy of the quantum optical treatment in such small scale since atoms' internal structures are not accounted for.

\section{Time evolution of De Moivre states}
To investigate the time evolution of De Moivre (DM) states, we turn to the Schr\"{o}dinger equations projected from the above Lindblad form.\ First we derive the time evolution of the singly-excited states $|\psi_\mu\rangle$ by projecting the coherence operators $\hat{Q}$ $=$ $|\psi_\mu\rangle\langle g|^{\otimes N}$ on $|g\rangle^{\otimes N}$.\ Define the state of the system in Schr\"{o}dinger picture as $|\Psi(t)\rangle$ $=$ $\sum_{\mu=1}^N c_\mu(t)|\psi_\mu\rangle$, we have
\bea
\dot{c}_\mu(t)=\sum_{\nu=1}^N M_{\mu\nu}c_\nu(t),
\eea
where $M_{\mu\nu}$ $\equiv$ $\frac{\Gamma}{2}(-F_{\mu\nu}+i2G_{\mu\nu}\delta_{\nu\neq\mu})$, forming a matrix $\hat{M}$ involving dynamical couplings between any pair of atoms.\ Using the similarity transformation, we can diagonalize $\hat{M}$ with the eigenvalues $\lambda_l$ and eigenvectors $\hat{U}$, such that
\bea
c_\mu(t)=\sum_{\nu,n} U_{\mu n}e^{\lambda_n t}U^{-1}_{n\nu}c_\nu(t=0),
\eea
where $c_\nu(t=0)$ denotes the initial condition of the system.\ When we prepare the atoms in one of the DM states $|\phi_m\rangle$, the state vector $|\Psi(t)\rangle$ can be expressed as $\sum_{m'=1}^N d_{m'}(t)|\phi_{m'}\rangle$, which starts to evolve initially from some DM state, that is $d_{m'}(t=0)$ $=$ $\delta_{m'm}$.\ Using the relation of $d_m$ $=$ $\sum_{\mu=1}^N$ $c_\mu e^{-i\k\cdot\r_\mu-i2m\pi(\mu-1)/N}/\sqrt{N}$, eventually we derive the time evolution of the DM states,
\bea
d_m(t)=\sum_{n=1}^N v_n(m) e^{\lambda_n t}w_n(m),
\eea
where 
\bea
v_n(m)&\equiv&\sum_{\mu=1}^N \frac{e^{-i\k\cdot\r_\mu-i2m\pi(\mu-1)/N}}{\sqrt{N}}U_{\mu n},\\
w_n(m)&\equiv&\sum_{\nu=1}^N U^{-1}_{n\nu}\frac{e^{i\k\cdot\r_\nu+i2m\pi(\nu-1)/N}}{\sqrt{N}}.
\eea
We note that $v_n(m)$ is the inner product of $m$th DM state and $n$th eigenvector in $\hat{U}$, in which $|v_n(m)|^2$ shows how close a DM state is to an eigen one.\ We also define a normalized weighting as $|v_n(m)w_n(m)|^2$ to describe the contribution of a specific $\lambda_n$ which governs the time evolution of the DM states.


\begin{thebibliography}{99}
\bibitem{Mandel1995} L. Mandel and E. Wolf, {\em Optical coherence and quantum optics}. (Cambridge University Press, 1995).
\bibitem{Hammerer2010} K. Hammerer, A. S. S{\o}rensen, and E. S. Polzik, Quantum interface between light and atomic ensembles. Rev. Mod. Phys. {\bf 82}, 1041 (2010).
\bibitem{Hartmann2006} M. J. Hartmann, F. G. S. L. Brand\~{a}o, and M P. Plenio, Strongly interacting polaritons in coupled arrays of cavities. Nat. Phys. {\bf 2}, 849 (2006).
\bibitem{Greentree2006} A. D. Greentree, C. Tahan, J. H. Cole, and L. C. L. Hollenberg, Quantum phase transitions of light. Nat. Phys. {\bf 2}, 856 (2006).
\bibitem{Lewenstein2007} M. Lewenstein, A. Sanpera, V. Ahufinger, B. Damski, A. Sen(De), and U. Sen, Ultracold atomic gases in optical lattices:
mimicking condensed matter physics and beyond. Adv. Phys. {\bf 56}, 243 (2007).
\bibitem{Chang2008} D. E. Chang, V. Gritsev, G. Morigi, V. Vuleti\'{c}, M. D. Lukin, and E. A. Demler, Crystallization of strongly interacting photons in a nonlinear optical fibre. Nat. Phys. {\bf 4}, 884 (2008).
\bibitem{Georgescu2014} I. M. Georgescu, S. Ashhab, and F. Nori, Quantum simulation. Rev. Mod. Phys. {\bf 86} 153 (2014).
\bibitem{Dicke1954} R. H. Dicke, Coherence in spontaneous radiation processes, Phys. Rev. {\bf 93}, 99 (1954).
\bibitem{Gross1982} M. Gross, and S. Haroche, Superradiance: An essay on the theory of collective spontaneous emission. Phys. Rep. {\bf 93}, 301 (1982).
\bibitem{Friedberg1973} R. Friedberg, S. R. Hartmann, and J. T. Manassah, Frequency shifts in emission and absorption by resonant systems of two-level atoms, Phys. Rep. {\bf 7}, 101 (1973).
\bibitem{Scully2009} M. O. Scully, Collective Lamb shift in single photon Dicke superradiance. Phys. Rev. Lett. {\bf 102}, 143601 (2009).
\bibitem{Scully2006} M. O. Scully, E. S. Fry, Ooi C. H. Raymond, and K. W\'{o}dkiewicz, Directed spontaneous emission from an extended ensemble of N atoms: Timing is everything. Phys. Rev. Lett. {\bf 96}, 010501 (2006).
\bibitem{Eberly2006} J. H. Eberly, Emission of one photon in an electric dipole transition of one among N atoms. J. Phys. B: At. Mol. Opt. Phys. {\bf 39}, S599 (2006).
\bibitem{Mazets2007} I. E. Mazets, and G. Kurizki, Multiatom cooperative emission following single-photon absorption: Dicke-state dynamics. J. Phys. B: At. Mol. Opt. Phys. {\bf 40}, F105 (2007).
\bibitem{Stephen1964} M. J. Stephen, First-order dispersion forces. J. Chem. Phys. {\bf 40}, 669 (1964).
\bibitem{Lehmberg1970} R. H. Lehmberg, Radiation from an N-atom system. I. General formalism, Phys. Rev. A {\bf 2}, 883 (1970).
\bibitem{Chaneliere2006} T. Chaneli\`{e}re, D. N. Matsukevich, S. D. Jenkins, T. A. B. Kennedy, M. S. Chapman, and A. Kuzmich, Quantum telecommunication based on atomic cascade transitions. Phys. Rev. Lett. {\bf 96}, 093604 (2006).
\bibitem{Srivathsan2013} B. Srivathsan, G. K. Gulati, B. Chng, G. Maslennikov, D. Matsukevich, and C. Kurtsiefer, Narrow band source of transform-limited photon pairs via four-wave mixing in a cold atomic ensemble. Phys. Rev. Lett. {\bf 111}, 123602 (2013).
\bibitem{Rohlsberger2010} R. R\"{o}hlsberger, K. Schlage, B. Sahoo, S. Couet, and R. R\"{u}ffer, Collective Lamb shift in single-photon superradiance. Science {\bf 328}, 1248-1251 (2010).
\bibitem{Keaveney2012} J. Keaveney, A. Sargsyan, U. Krohn, I. G. Hughes, D. Sarkisyan, and C. S. Adams, Cooperative Lamb shift in an atomic vapor layer of nanometer thickness. Phys. Rev. Lett. {\bf 108}, 173601 (2012).
\bibitem{Meir2014} Z. Meir, O. Schwartz, E. Shahmoon, D. Oron, and R. Ozeri, Cooperative Lamb shift in a mesoscopic atomic array. Phys. Rev. Lett. {\bf 113}, 193002 (2014).
\bibitem{Pellegrino2014} J. Pellegrino, R. Bourgain, S. Jennewein, Y. R. P. Sortais, A. Browaeys, S. D. Jenkins, and J. Ruostekoski, Observation of suppression of light scattering induced by dipole-dipole interactions in a cold-atom ensemble. Phys. Rev. Lett. {\bf 113}, 133602 (2014).
\bibitem{Sonnefraud2010} Y. Sonnefraud, N. Verellen, H. Sobhani, G. A.E. Vandenbosch, V. V. Moshchalkov, P. V. Dorpe, P. Nordlander, and S. A. Maier, Experimental realization of subradiant, superradiant, and Fano resonances in ring/disk plasmonic nanocavities. ACS Nano {\bf 4}, 1664 (2010)
\bibitem{McGuyer2015} B. H. McGuyer, M. McDonald, G. Z. Iwata, M. G. Tarallo, W. Skomorowski, R. Moszynski, and T. Zelevinsky, Precise study of asymptotic physics with subradiant ultracold molecules. Nat. Phys. {\bf 11}, 32 (2015).
\bibitem{Guerin2016} W. Guerin, M. O. Ara\'{u}jo, and R. Kaiser, Subradiance in a large cloud of cold atoms. Phys. Rev. Lett. {\bf 116}, 083601 (2016).
\bibitem{Jen2015} H. H. Jen, Superradiant cascade emissions in an atomic ensemble via four-wave mixing. Ann. of Phys. (N.Y.) {\bf 360}, 556 (2015).
\bibitem{Scully2015} M. O. Scully, Single photon subradiance: Quantum control of spontaneous emission and ultrafast readout. Phys. Rev. Lett. {\bf 115}, 243602 (2015).
\bibitem{Svidzinsky2008} A. A. Svidzinsky, J.-T. Chang, and M. O. Scully, Dynamical evolution of correlated spontaneous emission of a single photon from a uniformly excited cloud of N atoms. Phys. Rev. Lett. {\bf 100}, 160504 (2008).
\bibitem{Plankensteiner2015} D. Plankensteiner, L. Ostermann, H. Ritsch, and C. Genes, Selective protected state preparation of coupled dissipative quantum emitters, Sci. Rep. {\bf 5}, 16231 (2015).
\bibitem{Vetter2016} P. A. Vetter, L. Wang, D.-W. Wang, and M. O. Scully, Single photon subradiance and superradiance revisited: a group theoretic analysis of subradiant states. Physica Scripta {\bf 91}, 023007 (2016).
\bibitem{note1} On preparing this manuscript, we are aware of the relevant construction of the subradiant states \cite{Vetter2016} based on group theory.
\bibitem{Afzelius2015} M. Afzelius, N. Gisin , and H. Riedmatten, Quantum memory for photons. Physics Today {\bf 68}(12), 42 (2015).
\bibitem{Grimm2000} R. Grimm, M. Weidem\"{u}ller, and Y. B. Ovchinnikov, Optical dipole traps for neutral atoms, Adv. At. Mol. Opt. Phys. {\bf 42}, 95 (2000).
\bibitem{Sparkes2010} B. M. Sparkes, M. Hosseini, G. H\'{e}tet, P. K. Lam, and B. C. Buchler, ac Stark gradient echo memory in cold atoms, Phys. Rev. A {\bf 82}, 043847 (2010).
\bibitem{Steck2015} D. A. Steck, Rubidium 87 D line data, available online at http://steck.us/alkalidata (revision 2.1.5, 13 January 2015). 
\bibitem{Kraus2006} B. Kraus, W. Tittel, N. Gisin, M. Nilsson, S. Kroll, and J. I. Cirac, Quantum memory for nonstationary light fields based on controlled reversible inhomogeneous broadening. Phys. Rev. A {\bf 73}, 020302(R) (2006).
\bibitem{Hetet2008} G. H\'{e}tet, J. J. Longdell, A. L. Alexander, P. K. Lam, and M. J. Sellars, Electro-optic quantum memory for light using two-level atoms. Phys. Rev. Lett. {\bf 100}, 023601 (2008).
\bibitem{Hedges2010} M. P. Hedges, J. J. Longdell, Y. Li, and M. J. Sellars, Efficient quantum memory for light.  Nature {\bf 465}, 1052 (2010).
\bibitem{Hosseini2011} M. Hosseini, B. M. Sparkes, G. Campbell, P. K. Lam, and B. C. Buchler, High efficiency coherent optical memory with warm rubidium vapour. Nat. Comm. {\bf 2}:174 (2011).
\bibitem{Nienhuisi1987} G. Nienhuisi, and F. Schulleri, Spontaneous emission and light scattering by atomic lattice models. J. Phys. B: At. Mol. Phys. {\bf 20}, 23 (1987).
\bibitem{Paredes2004} B. Paredes, A. Widera, V. Murg, O. Mandel, S. F\"{o}ling, I. Cirac, G. V. Shlyapnikov, T. W. H\"{a}nsch, and I. Bloch, Tonks-Girardeau gas of ultracold atoms in an optical lattice. Nature {\bf 429}, 227 (2004).
\bibitem{Kinoshita2004} T. Kinoshita, T. Wenger, and D. S. Weiss, Observation of a one-dimensional Tonks-Girardeau gas. Science {\bf 305}, 1125 (2004).
\bibitem{Weimer2013} H. Weimer, N. Y. Yao, and M. D. Lukin, Collectively enhanced interactions in solid-state spin qubits. Phys. Rev. Lett. {\bf 110}, 067601 (2013).
\bibitem{Sipahigil2014} A. Sipahigil, K. D. Jahnke, L. J. Rogers, T. Teraji, J. Isoya, A. S. Zibrov, F. Jelezko, and M. D. Lukin, Indistinguishable photons from separated silicon-vacancy centers in diamond. Phys. Rev. Lett. {\bf 113}, 113602 (2014)
\bibitem{Chang2012} D. E. Chang, L. Jiang, A. V. Gorshkov, and H. J. Kimble, Cavity QED with atomic mirrors. New J. Phys. {\bf 14}, 063003 (2012).
\bibitem{Olmos2013} B. Olmos, D. Yu, Y. Singh, F. Schreck, K. Bongs, and I. Lesanovsky, Long-range interacting many-body systems with alkaline-earth-metal atoms. Phys. Rev. Lett. {\bf 110}, 143602 (2013).
\bibitem{Bettles2015} R. J. Bettles, S. A. Gardiner, and C. S. Adams, Phys. Rev. A {\bf 92}, 063822 (2015).
\bibitem{Saffman2010} M. Saffman, T. G. Walker, and K. M\o lmer, Quantum information with Rydberg atoms. Rev. Mod. Phys. {\bf 82}, 2313 (2010).
\bibitem{Peyronel2012} T. Peyronel, O. Firstenberg, Q. Liang, S. Hofferberth, A. V. Gorshkov, T. Pohl, M. D. Lukin, and V. Vuleti\'{c}, Quantum nonlinear optics with single photons enabled by strongly interacting atoms. Nature {\bf 488}, 57 (2012).
\end{thebibliography}
\end{document}